\documentclass[cits]{PoS}

\title{Two Color Matter in the Quenched Approximation}

\ShortTitle{Two Color Matter in the Quenched Approximation}

\author{\speaker{Pietro Giudice} \\
        Dipartimento di Fisica Teorica, Università di Torino and INFN, sezione di Torino, Italy\\
        E-mail: \email{giudice@to.infn.it}}

\author{Simon Hands \\
        Physics Department, Swansea University, Singleton Park, Swansea SA2 8PP, United Kingdom \\
        E-mail: \email{s.hands@swan.ac.uk}}

\abstract{
We study a quenched SU(2) lattice gauge theory in 4d in which 
the spatial gauge ensemble $\{U_i\}$ is generated 
from a 3d gauge-Higgs model and the
timelike link variables are ``reconstructed'' from the Higgs fields. 
The resulting ensemble is used to study quenched quark propagation 
with non-zero chemical potential $\mu$. 
While it proves possible to alter the strength of the inter-quark 
interaction by changing the parameters of the dimensionally reduced model, 
there is no evidence for any region of parameter space where quarks 
exhibit deconfined behaviour or thermodynamic observables scale as if 
there were a Fermi surface.
}

\FullConference{The XXV International Symposium on Lattice Field Theory\\
		 July 30 - August 4 2007\\
		 Regensburg, Germany}

\newcommand{\bi}  {\begin{itemize}}
\newcommand{\ei}  {\end{itemize}}
\newcommand{\be}  {\begin{enumerate}}
\newcommand{\ee}  {\end{enumerate}}

\newcommand{\bc}  {\begin{center}}
\newcommand{\ec}  {\end{center}}

\newcommand{\eqa}{\begin{eqnarray}}
\newcommand{\ena}{\end{eqnarray}}

\newcommand{\eq}{\begin{equation}}
\newcommand{\en}{\end{equation}}
\begin{document}

\section{Introduction}

Lattice QCD suffers from the so-called Sign Problem;
the Lagrangian density with $N_f$ quark flavors has the form $\bar q Mq$; 
the functional measure $\mbox{det}^{N_f}M(\mu)=\mbox{det}^{N_f}M^*(-\mu)$ 
implies that for $\mu\not=0$ the action is complex, rendering 
Monte Carlo importance sampling impracticable.

Can one at least perform lattice QCD simulations in the quenched 
$N_f\to0$ limit, i.e. study the propagation of valence quarks 
with $\mu\not=0$ through a non-perturbative gluon 
background~\cite{Giudice:2007yv}?

Of course, the information extracted from such an approach could be at best
qualitative, since (unlike the case of $T>0$) the gauge field ensemble 
$\{U_\mu\}$ can only respond to $\mu\not=0$ via virtual quark loops.
Anyway, such information might be valuable in furnishing a non-perturbative 
definition of the Fermi surface, whose existence is assumed in most 
phenomenological treatments of dense matter.

In the context of a random matrix theory,
Stephanov~\cite{Misha} showed that the quenched theory should be thought of 
as the $N_f\to0$ limit of a QCD-like theory with not just $N_f$ flavors of 
quark $q\in{\bf3}$ of the SU(3) gauge group 
but also with $N_f$ flavors of conjugate quark $q^c\in\bar{\bf3}$.
As a consequence, $qq^c$ bound states appear in the  spectrum, 
resulting in baryons degenerate with light mesons.
For $\mu/T\gg1$ there is an {\em onset transition} from the vacuum to a 
ground state with quark number density $n_q~>0~$.
For QCD it occurs at $\mu_o\simeq m_N/3$ ($m_N$ is the nucleon mass) but if 
conjugate quarks are present   $\mu_o\simeq m_\pi/2$.

We try to modify the gluon background in some way so that color confinement 
no longer holds:  then the role of $qq^c$
excitations may not be so important in determining the ground state in the 
quark sector.
Our hope is that valence quark propagation in such a background may 
qualitatively resemble that of the deconfined regime of the phase diagram 
at $\mu/T\gg1$.
We start with the 3$d$ configurations characteristic 
of the deconfined phase found at $T>T_c$, $\mu/T\ll1$
produced by the approach to hot gauge theory known as 
Dimensional Reduction (DR).
The quenched action is the 3$d$ SU(2) gauge -- adjoint Higgs model 
obtained by DR from 4$d$ SU(2) given by
Eqn. (4) of Ref.\cite{HP1}:
$$
S_{3d}=\beta\sum_{x,i>j}\left(1-{1\over2}\mbox{tr}U_{x,ij}\right)+
2\sum_x\mbox{tr}(\varphi_x\varphi_x)
-2\kappa\sum_{x,i}\mbox{tr}(\varphi_x
U_{x,i}\varphi_{x+\hat\imath}U_{x,i}^\dagger)+
\lambda\sum_x(2\mbox{tr}(\varphi_x\varphi_x) - 1)^2,
$$
where $\varphi\equiv{1\over2}\varphi_a\tau_a$ 
represents the adjoint Higgs field.
In the DR approach, all non-static modes of the gauge theory
are integrated out leaving a 3$d$ gauge-Higgs model describing the 
non-perturbative behaviour of the remaining static modes.

Our goal is to study quark propagation through a 
{\em non-confining\/} quenched gluon background with $\mu\not=0$. Since
chemical potential couples to quarks via the timelike component of the 
current $\mu\bar\psi\gamma_0\psi$, this is an
inherently four-dimensional problem. In order to 
generate such a background we take a 3$d$ configuration generated
by the DR simulation, 
motivated by the fact that it describes deconfining physics, and
``reconstruct'' the gauge field in the timelike direction 
via the prescription:
\begin{equation}
U_0=
\exp\left(ig\sqrt{\kappa\over
N_\tau}\varphi\right)
=\cos\left(\tilde g\sqrt{\varphi_a\varphi_a}\right)
+i{{\tau_a\varphi_a}\over\sqrt{\varphi_a\varphi_a}}
\sin\left(\tilde g\sqrt{\varphi_a\varphi_a}\right),
\label{eq:extend}
\end{equation}
with $\tilde g=\sqrt{\kappa\over\beta}$.
Spatial link variables $U_i$ are taken to be time independent and identical 
to their 3$d$ counterparts.
Henceforth we use 4$d$ configurations $\{U_\mu\}$ generated 
as outlined above as input in quenched studies.

Since $q$ and $\bar q$ fall in equivalent
representations of the gauge group hadron multiplets contain both $q\bar q$
mesons and $qq$, $\bar q\bar q$ baryons, which are degenerate at $\mu=0$. 
In the chiral limit the lightest hadrons are Goldstone bosons and 
can be analysed using chiral perturbation theory ($\chi$PT) \cite{KSTVZ}:
at leading order for $\mu\gg T$ a second order 
onset transition from vacuum to matter consisting of tightly bound diquark
scalar bosons is predicted at exactly $\mu_o=m_\pi/2$. 
In the limit $\mu\to\mu_{o+}$ the
matter in the ground state becomes arbitrarily dilute, weakly-interacting and
non-relativistic: a textbook example of Bose-Einstein condensation.
At the same point the chiral condensate
$\langle\bar qq\rangle$ starts to fall below its vacuum value and a
non-vanishing diquark condensate $\langle qq\rangle$ develops.
The diquark condensate spontaneously breaks U(1) baryon number symmetry, so
the ground state is superfluid.

More recent simulations have found evidence for a second transition at larger
$\mu$ to a deconfined phase, as evidenced by a non-vanishing Polyakov
loop~\cite{HKS} and by a fall in the topological susceptibility~\cite{ADL}.
In this regime thermodynamic quantities scale according to the expectations of 
free field theory (also referred to as ``Stefan-Boltzmann'' (SB) scaling), 
namely $n_q\propto\mu^3$, and energy density 
$\varepsilon\propto\mu^4$~\cite{HKS}.

\section{Numerical Results}

\begin{figure}
\begin{center}
\includegraphics[width=6.5cm]{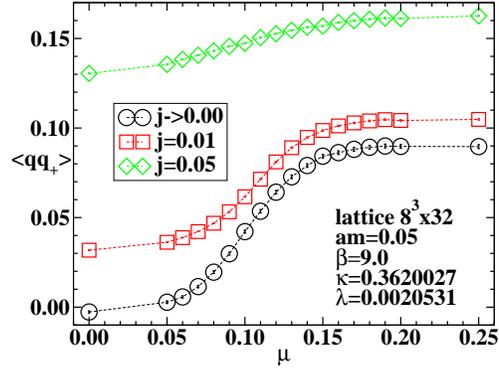}
\caption{$\langle qq_+\rangle$ vs. $\mu a_t$ for various $j$; 
the extrapolation to $j\to0$ is obtained using a cubic polynomial.}
\label{fig:qqvsmu_kappa0.36}
\end{center}
\end{figure}

We have chosen $\beta=9.0$ sufficiently close to the continuum limit for the 
DR formalism to be trustworthy, and start with a point with $T=2T_c$:
this corresponds to $\kappa=0.3620027$, $\lambda=0.0020531$.
It is important to make a precise determination of the pion mass at $\mu=0$:
we obtain $m_\pi a_t=0.2321(1)$ on a $8^3 \times 32$. 
Since this scale is not too dissimilar to
$L_s^{-1}$, we have repeated the measurement on $16^3\times64$, where we find
$m_\pi a_t=0.2368(3)$. The systematic error due to finite volume is 
significant, but small enough at 2\% to be acceptable for this exploratory 
study.
We have analysed the quark density $n_q$ and chiral condensate 
$\langle\bar qq\rangle$ as functions of $\mu$ for various $j$ with
$0\leq\mu/T\leq8$.
There is a transition at $\mu a_t\approx0.12$, 
becoming more abrupt as 
diquark source
$j\to0$, from a phase with $n_q=0$, $\langle\bar qq\rangle$ constant to 
one in which $n_q$ increases approximately linearly with $\mu$ and
$\langle\bar qq\rangle\propto\mu^{-2}$. 
This is in complete accordance with the
scenario described by $\chi$PT in which as $\mu$ increases at $T\approx0$
there is a transition at $\mu_c=m_\pi/2$ from the vacuum to a 
weakly-interacting Bose gas formed from scalar diquarks.
The diquarks are supposed to Bose-condense to form a superfluid condensate;
on a finite system this must be checked at $j\not=0$ using the 
$\langle qq_+\rangle$ observable:
the condensate increasing monotonically with $\mu$.
To determine the nature of the ground state
an extrapolation $j\to0$ is needed. We have used
a cubic polynomial for data with $0.02\leq ja\leq0.1$; 
Fig.~\ref{fig:qqvsmu_kappa0.36} 
confirms that once again there is an abrupt 
change of behaviour in the order parameter at $\mu\approx m_\pi/2$, and
that the high-$\mu$ phase is superfluid.
Next we explored a parameter set corresponding to a smaller scalar
``stiffness'' by changing to $\kappa=0.1$. Naively this corresponds to a 
huge value of $T/T_c$, ie. taking us further into
the deconfined phase of the hot gauge theory.  Of course, whether 
DR-based concepts
remain valid for the reconstructed theory must be addressed empirically.
This time we used a volume $8^3\times64$ for the bulk observables
and at $\mu=0$ determined the pion mass $m_\pi a_t=0.1377(1)$ on $8^3\times
128$, and $m_\pi a_t=0.1423(4)$ on $16^3\times64$, 
showing that the finite volume error is now roughly 3\%.
We studied $n_q$ and $\langle\bar qq\rangle$ as functions of $\mu$ 
for $0 \leq \mu / T \approx 12$. It is noteworthy that for $\mu>m_\pi/2$
$n_q(\mu)$ is numerically very similar to the values found at 
$\kappa=0.3620027$, whereas for $\mu<m_\pi/2$
the chiral condensate $\langle\bar qq\rangle$ is significantly smaller,
indicative of a weaker quark -- anti-quark binding at this smaller $\kappa$.
Moreover, using the Gell-Mann-Oakes-Renner relation, valid for $\chi$PT,  
$f_\pi^2 m_\pi^2 = 2 m_q \langle\bar qq\rangle$ we have:  
$f_\pi^2 a^2=0.99$ for $\kappa=0.1$ and $f_\pi^2 a^2=0.58$ for 
$\kappa=0.3620027$. 
The non-interacting limit is $f_\pi \to \infty$ therefore the 
interparticle intractions are weaker at $\kappa=0.1$.
As before, there is a clear discontinuity in the
observables' behaviour at $\mu_c\simeq m_\pi/2$, and the general picture is
qualitatively very similar, suggesting that the $\chi$PT
scenario is still applicable. Diquark binding is now also  much weaker.

It is disappointing that we have found no qualitative change in physics as the
parameters are varied -- recall that the $\chi$PT model which describes the
results reasonably well is based on the assumption of confinement, or at least
on the presence of very tightly bound diquark states in the spectrum.

To explore the parameter space more widely we focussed on a single observable,
$n_q$, and scanned the $(\kappa,\lambda)$ plane on $8^3\times16$
at five different values of
$\mu$ with $\beta=9.0$, $ma=0.05$ and $ja=0.01$.
Data shows that except for $\lambda=0.1$ the results for fixed $\mu$ 
are practically independent of $\kappa$ and of $\lambda$. 
Moreover, $n_q$ increases linearly with $\mu$ over a wide region of 
parameter space, as it does for $\mu>\mu_c$. 
This approximate linear behaviour is once again a prediction of $\chi$PT, 
and is to be contrasted with the $n_q\propto\mu^3$ behaviour
expected of a deconfined theory where baryons can be identified with 
degenerate quark states occupying a Fermi sphere of radius $k_F\approx\mu$.
The absence of this scaling is a further reason to conclude that the
reconstructed model does not describe deconfined physics.

\subsection{Bosonic Spectrum}

An appropriate set of states to look at, to study the bosonic spectrum, 
is constituted of pion, scalar, higgs and goldstone.
Pion and scalar states are related via the
U(1)$_\varepsilon$ global symmetry $\chi\mapsto e^{i\alpha\varepsilon}\chi$,
$\bar\chi\mapsto\bar\chi e^{i\alpha\varepsilon}$. Analogous to
chiral symmetry for continuum spinors, this is an exact symmetry of the 
action in the limit $m\to0$. 
In a phase with spontaneously
broken chiral symmetry, the scalar is massive, and the pion a Goldstone mode,
becoming massless as $m\to0$. Similarly, ``higgs'' and ``goldstone'' diquark 
states are related via the U(1)$_B$ baryon number rotation
$\chi\mapsto e^{i\beta}\chi$, $\bar\chi\mapsto\bar\chi e^{-i\beta}$, an exact
symmetry of the action in the limit $j\to0$.
In a superfluid phase with $\langle qq_+\rangle\not=0$, the higgs is massive,
and the goldstone massless in the limit $j\to0$.

The boson correlators are constructed from the {\em Gor'kov} 
propagator where appear the $2\times2$ (in color space) matrices 
$N\sim\langle\chi_x\bar\chi_y\rangle$ and $A
\sim\langle\chi_x\chi_y\rangle$ which are known as the {\em normal} and {\em
anomalous} parts respectively. 
On a finite volume $A\equiv0$ for $j=0$; 
$\lim_{j\to0}\lim_{V\to\infty}A\not=0$ signals particle-hole mixing resulting
from the breakdown of U(1)$_B$ symmetry, and hence superfluidity.
Due to SU(2) symmetries the only 
independent components of ${\cal G}$ are $\mbox{Re}N_{11}\equiv N$ and 
$\mbox{Im}A_{12}\equiv A$ and their barred counterparts.

We have studied the model with a chemical potential $\mu a_t=0.25$, 
ie. above the critical $\mu_c$ required to enter the superfluid phase. 
All four channels yield clear signals for single particle bound states.

Like any meson constructed from staggered fermions,
the correlators in principle describe two states and must
be fitted using the form
\begin{equation}
C(t)=A[e^{-mt}+e^{-m(L_t-t)}]+B[e^{-Mt}+(-1)^te^{-M(L_t-t)}],
\label{eq:propfit}
\end{equation}
where $m$ and $M$ denote the masses of states with opposite parities.
In most cases we find $M\gg m$; however for $\mu>\mu_c$ the pion correlator 
has a distinct ``saw-tooth'' shape, and in fact the fit yields
$m_\pi>M_{b1}$, where $\pi$ denotes the usual pseudoscalar pion, and $b1$ a
state of opposite parity, which must therefore be scalar.
\begin{figure}
\begin{center}
\includegraphics[width=6.5cm]{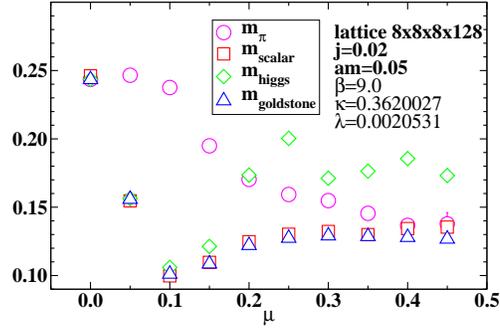}
\caption{Mass spectrum of various bosonic excitations as a function of 
$\mu a_t$.}
\label{fig:bosons}
\end{center}
\end{figure}
In Fig.~\ref{fig:bosons} the corresponding spectrum for all the boson states.  
Note that all states are approximately degenerate at $\mu=0$; 
the equality of pion, higgs and goldstone correlators 
is guaranteed by SU(2) symmetry at $\mu=0$, but the degeneracy of
the scalar in the chirally-broken vacuum can only arise as a result 
of meson-diquark mixing due to $j\not=0$.
Next, note that the pion mass remains constant for $\mu<\mu_c$, where 
it is a pseudo-Goldstone boson associated with chiral symmetry breaking, 
and then falls once the superfluid phase is entered;
this is in accordance with the predictions of $\chi$PT. 
Most of the other states show a much steeper decrease with $\mu$ for 
$\mu<\mu_c$, followed by a gentle rise to a
plateau at $ma\approx0.13$ in the superfluid phase $\mu>\mu_c$, 
precisely as expected of the goldstone state in the superfluid 
phase with diquark source $j\not=0$.

The exception is the higgs, which rises more steeply to become the 
heaviest state at large $\mu$.  
We conclude that the breaking of degeneracy between higgs
and goldstone states is clear
supplementary evidence for the breaking of U(1)$_B$ symmetry in the superfluid
phase and all states with $J^P=0^+$ including the $b1$ but {\em
except\/} the higgs have some projection onto the Goldstone state, 
regardless of whether the original interpolating operator is mesonic 
or baryonic.

\subsection{Fermionic Spectrum}

We have also studied the fermion spectrum, often in the context of 
condensed matter called the {\em quasiparticle} spectrum.
Since the Gor'kov propagator ${\cal G}$ is not gauge invariant we have 
to specify a gauge fixing procedure. 
A feature of the quenched approach is that it permits large statistics to be
accumulated with relatively little CPU effort. This has enabled us for the
first time in a gauge theory context to study ${\cal G}$ at
$\mu\not=0$, by helping to overcome the sampling problems associated with 
gauge fixing.
We have experimented with two gauge choices:
{\em Unitary gauge\/} $\varphi\mapsto\varphi^\prime=
(0,0,\varphi_3^\prime)$, which is implemented {\em before} 
the reconstruction of the
4th dimension, and is unique up to a Z$_2$ factor, specified
by demanding $\varphi^\prime_3\geq0$; 
and {\em Coulomb gauge\/}, implemented by 
maximising $\sum_{xi}\mbox{tr}(U_{x,i}+U^\dagger_{x-\hat\imath,i})$, 
in an attempt to make the gauge fields as smooth as possible and
hence improve the signal-to-noise ratio.
We have studied the normal and anomalous fermion timeslice propagators on a
$32\times8^2\times64$ lattice at $\beta=9.0$, $\kappa=0.1$, 
$\lambda=0.0020531$, $ma=0.05$, $ja=0.0.2$
and $\mu a_t=0.3$; the last value chosen to ensure $\mu>\mu_c$. 
Two features to note are that the 
Coulomb data is roughly twice as large as the unitary data reflecting an
enhanced signal and that there is little variation with $k_x$. 

In the NJL model the quasiparticle propagator can be successfully 
fitted using the forms 
\begin{eqnarray}
N(t)=Pe^{-E_Nt}+Qe^{-E_N(L_t-t)},\label{eq:fitN}\\
A(t)=R[e^{-E_At}-e^{-E_A(L_t-t)}],\label{eq:fitA}
\end{eqnarray}
where for $\mu\not=0$ there is no reason to expect $P=Q$, but for a 
well-defined quasiparticle state the equality $E_N=E_A$ should hold.
Here, by contrast, only the anomalous channel fits produced an 
acceptable $\chi^2$ and shows any evidence of gauge independence. 
The value of $E_A$ obtained is very close to
$m_\pi/2$, indicating that at this value of $\kappa$ the pion is a weakly bound
state.
Another striking feature of the data is the
approximate forwards-backwards symmetry of $N(t)$, implying $P\simeq Q$. 
We have studied the dispersion relations $E_A(k_x)$ for data
taken on a $32\times8^2\times64$ lattice;
it confirms that the quasiparticle excitation energies are $k$-independent:
one motivation for
our study was to investigate to what extent the concept of a Fermi surface,
which is not strictly
gauge invariant, can be put on a firm empirical footing in a gauge
theory. Our results shows no evidence for a Fermi surface.
\begin{figure}
\begin{minipage}{70mm}
\begin{center}
\includegraphics[width=6.5cm]{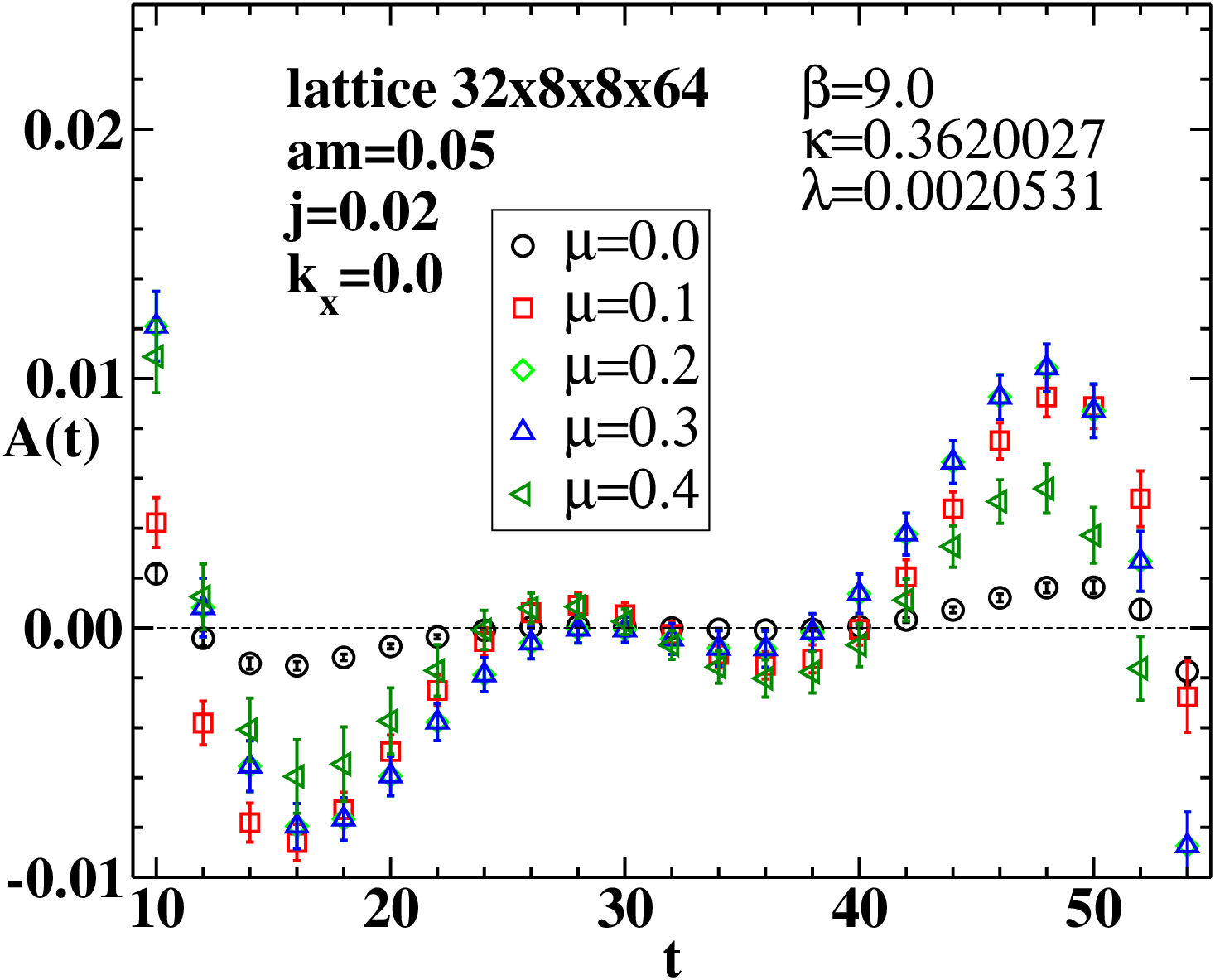}
\caption{Close-up of $A(t)$for various $\mu a_t$.}
\label{fig:wiggles}
\end{center}
\end{minipage}
\begin{minipage}{70mm}
\begin{center}
\includegraphics[width=6.5cm]{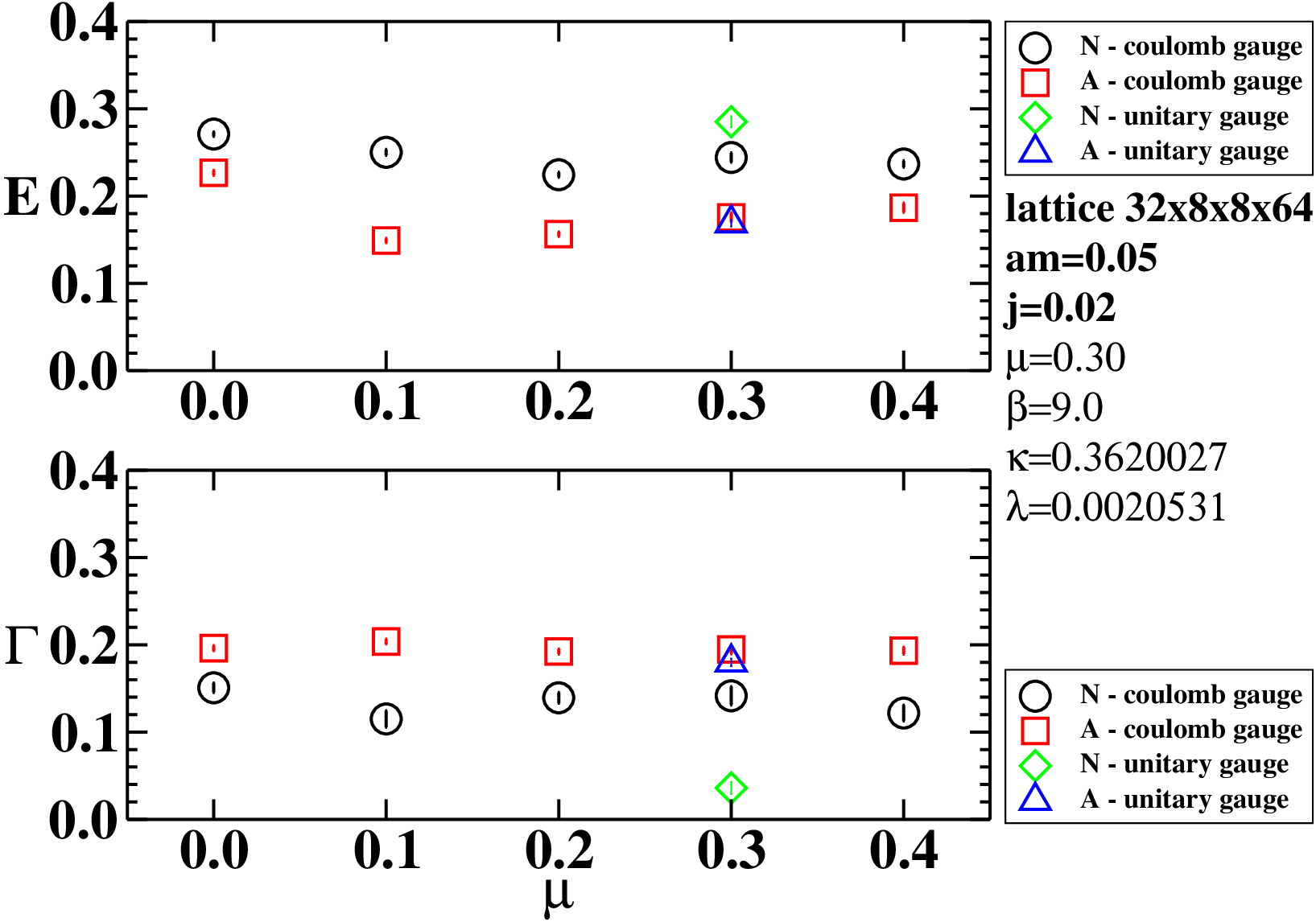}
\caption{$E$ and $\Gamma$ vs. $\mu a_t$ in both normal and 
anomalous channels.}
\end{center}
\end{minipage}
\end{figure}
The nature of the quasiparticle excitation is clarified a little
at the other parameter set studied, namely $\kappa=0.3620027$. 
In this case our results show no evidence for any well-defined spin-${1\over2}$
state in either normal or anomalous channels; as a result of confinement
the excitation spectrum of the model seems to be saturated by the tightly
bound spin-0 states of Fig.~\ref{fig:bosons}.
Fig.~\ref{fig:wiggles}
shows a close-up of $A(t)$ for various $\mu$ values, showing the presence of an
oscillatory component whose amplitude initially grows with $\mu$ 
(the $\mu a_t=0.3$
points overlay those from $\mu a_t=0.2$), but whose wavelength is roughly
$\mu$-independent. The origin of the oscillation could possibly be associated
with the non-unitarity of the model, but is most
likely a manifestation of independent spin-${1\over2}$ excitations being
ill-defined due to confinement. 

We have fitted the $\kappa=0.3620027$ data to the forms
\begin{eqnarray}
N(t)=Pe^{-E_Nt}\cos(\Gamma_Nt+\phi)+Qe^{-E_N(L_t-t)}\cos(\Gamma_N(L_t-t)+\phi)
,\label{eq:fitNosc}\\
A(t)=R[e^{-E_At}\cos(\Gamma_At+\phi)
-e^{-E_A(L_t-t)}\cos(\Gamma_A(L_t-t)+\phi)],\label{eq:fitAosc}
\end{eqnarray}
where we interpret $E$ as the energy and $\Gamma$ as the width of a
quasiparticle excitation.
The results' most striking feature is their independence
of $\mu$, with $\Gamma$ of the same order of magnitude as $E$. 
An interesting systematic
effect is that $E_N>E_A$ while $\Gamma_N<\Gamma_A$, which has motivated us 
to study $\sqrt{E^2+\Gamma^2}$ vs. $\mu$: 
the disparity between normal and anomalous channels is significantly reduced. 
Inspection of the $\mu a_t=0.3$ data also shows that the gauge 
dependence of this result is O(20\%) at worst. 
Numerically, $\sqrt{E^2+\Gamma^2}>m_\pi$, indicating strong 
quark -- anti-quark binding, due to the persistence of confinement at this 
value of $\kappa$.
Therefore we can interpret the effect of confinement as rotating
the quasiparticle pole into the complex plane, the rotation angle being larger
in the anomalous channel than in the normal one.

\section{Conclusions}

Our attempt to alter the nature of the gluon background by changing the
parameters of the 3$d$ DR gauge-Higgs model has been a partial success, in that
in going from $\kappa=0.3620027$ to $\kappa=0.1$ 
the strength of the binding between quarks weakens significantly. 

However, in neither case is there evidence for significant departure
of $n_q$, $\langle\bar qq\rangle$ and $\langle qq\rangle$ from the behaviour 
predicted by $\chi$PT, so that even if quarks are important degrees of 
freedom at $\kappa=0.1$, there is no evidence for the formation of
a degenerate system signalled by SB scaling. 

Sadly though, it appears to remain
the case that despite its ``unreasonable effectiveness'' in virtually
all other aspects of lattice QCD, the quenched
approximation has nothing useful to tell us about the physics of high quark
density.

\end{document}